\documentclass[iop,apjl]{emulateapj}
\usepackage{apjfonts}

\newcommand{\kms}{km~s$^{-1}$}
\newcommand{\halpha}{H$\alpha$\,}
\newcommand{\OI}{[OI]$\lambda$\,6300}
\newcommand{\Msun}{M$_{\odot}$}

\shorttitle{Bipolar Jets from a Spectroscopic Binary}
\shortauthors{Mundt et al.}

\begin{document}


\title{Bipolar Jets Produced by a Spectroscopic Binary
}


\author{Reinhard Mundt\altaffilmark{}}
\affil{Max Planck Institute for Astronomy, D-69117 Heidelberg,
   Germany}
\email{mundt@mpia.de}

\author{Catrina M. Hamilton\altaffilmark{}}
\affil{Physics and Astronomy Department, Dickinson College, 
       Carlisle, PA 17013}

\author{William Herbst \altaffilmark{}}
\affil{Van Vleck Observatory, Wesleyan University, 
      Middletown, CT 06459}

\author{Christopher M. Johns-Krull \altaffilmark{}}
\affil{Department of Physics and Astronomy, Rice University, Houston, 
      TX 77005}

\author{Joshua N. Winn \altaffilmark{}}
\affil{Department of Physics, and Kavli Institute for Astrophysics and Space Research,\\Massachusetts Institute of Technology, 
      77 Massachusetts Avenue, Cambridge, MA 02139}


\altaffiltext{}{}


\begin{abstract}
  We present evidence that the spectroscopically identified bipolar
  jets of the pre-main sequence binary KH\,15D (P=48.4\,d,
  $\epsilon\sim$0.6, periastron separation $\sim$18\,R$_A$,
  M$_A$=0.6\Msun , M$_B$=0.7\Msun ) are a common product of the whole
  binary system, rather than being launched from either star
  individually. They may be launched from the innermost part of the
  circumbinary disk (CBD) or may result from the merging of two
  outflows driven by the individual stars.  This evidence is based on
  high-resolution \halpha\ and \OI\ line profiles obtained during
  eclipse phases of this nearly edge-on system.  The occultation of
  star A (the only currently visible star) by the disk strongly
  suppresses the stellar \halpha\ and continuum emission and allows
  one to study the faint redshifted and blueshifted emission
  components of the bipolar jets.  The strongest evidence for jet
  production by the whole binary system comes from the observed radial
  velocity symmetry of the two jet components relative to the systemic
  velocity of the binary in combination with current accretion models
  from the CBD onto a binary system.
\end{abstract}





\section{Introduction}

The models proposed for driving jets and outflows from young stars can
be roughly categorized as disk-driven outflows (e.g.\ Pudritz et
al. 2007), X-winds (e.g.\ Shang et al. 2007), and stellar wind-driven
flows (e.g.\ Matt \& Pudritz 2008). In all of these models the young
star is surrounded by a circumstellar disk with an outer radius of
about 10 - 100\,AU and an inner radius of only a few stellar radii. So
far it has not been investigated whether any of these models can be
applied to a close binary system such as KH\,15D, where the
spectroscopic signature of a bipolar outflow has been found in the
[OI]$\lambda$6300 line (Hamilton et al.\ 2003).  A small-scale bipolar
jet of $\sim$70\,AU length in [SII]$\lambda$6731 is also known for the
close 19\,d TTS binary UZ Tau\,E (Hirth et al. 1997) and the
[OI]$\lambda$6300 line profile of the 15.8\,d TTS binary DQ Tau of
Huerta et al.\ (2005) suggest that this system also drives an
outflow. In this Letter we discuss for the first time simple possible
models for jet formation in close TTS binary systems and hope that our
Letter will initiate detailed theoretical and observational studies of
this challenging astrophysical problem.

In recent years KH\,15D (age$\sim$2\,Myr, d=760\,pc) has been the
subject of many detailed observational and theoretical studies
(e.g. Winn et al.\ 2006; Herbst et al.  2008 and references
therein). It first got attention for its unique photometric
variability with deep and practically grey eclipses every 48.4 days
that have become deeper and wider with each passing year (Hamilton et
al.\ 2005; LeDuc et al.\ 2010).  As discussed in these references the
following model describes the available data.  A nearly edge-on binary
system is surrounded by a circumbinary disk (CBD) which is inclined to
the binary plane by $\approx$10-20$^{\circ}$ (Winn et al.\ 2004;
Chiang \& Murray-Clay 2004).  Since more than a decade the CBD has
been totally occulting the orbit of star B and has been increasingly
occulting that of star A.  This is due to its precession on a
$\sim$10$^3$\,yr time scale, which causes the occulting edge to cover
the orbit of the binary. During the epoch of observations discussed
here, only eclipses of star A are observed, which result from the
disappearance of this star behind the occulting edge.  Without this
sharp edge acting as a ``natural coronagraph'' it would have been
practically impossible to trace in \halpha\ or [OI] the much fainter
emission from the outflowing gas analyzed here.\\

\section{Observations}

The high-resolution spectra analyzed here have been obtained during
various observing runs with UVES at the VLT (R$\sim$50000, slit width
0.8-1.0$''$) and with HIRES at Keck (R$\sim$70000, slit width
0.8$''$). A full description of the data and their reduction wiil be
given in Hamilton et al.\ (2010).  At the distance of KH\,15D, 1.0$''$
corresponds to 760\,AU.  The observing dates of the individual spectra
are indicated in Figs. 1 and 2. The continuum levels of all these
spectra have been calibrated onto a common flux scale by using the
Cousins I band fluxes that were measured simultaneously (or nearly so)
in our CCD imaging campaigns.  One uncertainty in this procedure is
the system's R-I color, since R band fluxes are requested for
calibration of \halpha\ and [OI]. The R-I color was calculated on the
basis of the 2002/2003 data set (Hamilton et al.\ 2005). In the
spectra presented here, any uncertainty will most probably affect only
the two spectra taken at minimum light (i.e. from Dec. 14, 2001 and
March 12, 2004), since only there is the R-I versus I correlation
unreliable.






\section{Jet properties derived from [OI]$\lambda$\,6300 and \halpha\ }

It is now generally accepted that the forbidden emission lines (FELs)
of T Tauri stars (TTSs) are an important diagnostic for their outflows
close to the star on scales of 20-40\,AU (Hirth et al.\ 1997; Hartigan
et al.\ 1995; Eisl\"offel et al.\ 2000).  The most important line for
these studies is the [OI]$\lambda$\,6300 line, which is usually the
strongest FEL in TTSs in the optical and therefore most easily
observable. For KH\,15D, the rather small measured out of eclipse
W$_{\lambda}$(\OI )=0.4-0.5\,\AA\ (which would be $\sim$0.2\,\AA\ if
both stars would be unocculted), compared to classical TTS (CTTS) jet
sources, indicates a mass outflow rate that is $\sim$10 times smaller
than in other CTTSs (e.g. Hartigan et al.\ 1995).

In principle the \halpha\ line is a much better diagnostic for
investigating outflows on very small scales (e.g. 0.1-10\,AU from the
source) because it is not quenched above a certain electron density
like FELs. Note that the critical density of the \OI\ line of
$\approx10^6$\,$cm^{-3}$ is among the highest of all FELs observable
in the red spectral region of TTSs.  However, in most TTSs the very
bright stellar (magnetospheric) \halpha\ emission makes studies close
to the star extremely difficult.  For KH\,15D, however the natural
coronagraphic mask (see Sect. 1) strongly suppresses the stellar
\halpha\ emission and allows to study the faint \halpha\ emission from
the outflow.

In Fig. 1 we display all our nine UVES/VLT \OI\ line profiles.  They
were all taken with UVES since the HIRES spectra do not cover this
line.  In the top four ``uneclipsed'' or ``partially eclipsed''
profiles the superimposed stellar absorption spectrum has been
subtracted to minimize it's effects on the profile shape (see caption
for details).  The three 2001 profiles have already been published by
Hamilton et al.\ (2003).  In Fig. 2 we show 5 \halpha\ line profiles
obtained with UVES and HIRES on various nights during eclipse, which
are quite different compared to ``uneclipsed'' profiles (Hamilton et
al.\ 2003, 2010).  In both figures the given velocity scale is
relative to the systemic velocity (v${_{hel}}$ = 18.6\,\kms ).

The nine [OI] profiles displayed in Fig. 1 are very similar and only
small variations are indicated.  In consideration of the night sky and
flux calibration uncertainties, and the additional noise in the top
four profiles due to the absorption spectrum subtraction, it is
unclear, whether the v$_{rad}$ values and the relative fluxes (f)
shown in Fig. 1 indicate any significant variation.  The only
exception is the profile from Dec. 20, 2001, where the flux tropped by
about a factor of 2 (see Sect. 5 for an interpretation).  In most
profiles in Fig. 1 two distinct (but partly blended) peaks at
$\sim$-22\,\kms\ and $\sim$+25\,\kms\ are visible.  On the basis of
the three 2001 [OI] profiles Hamilton et al.\ (2003) interpreted them
in terms of a bipolar jet in which the two jet components lie close to
the plane of the sky resulting in relatively small radial
velocities. The \halpha\ line profiles obtained during eclipse
strengthen this interpretation, since several of them show signatures
of bipolar jets similar to the [OI] line profile.  A particularly nice
example of a double-peaked profile is the middle one in Fig. 2 from
Dec. 16, 2003.  When comparing the [OI] line profiles with the
\halpha\ line profiles it is evident that the former indicate lower
velocities for the bipolar jets.  This could be explained by the
presence of a blended and unresolved low-velocity component, which is
often observed in the FELs of CTTSs, but not in \halpha.

Compared to [OI] the presented \halpha\ line profiles indicate
considerable variability, particularly in the red jet component (which
is often much fainter than the blue one) and in the broad \halpha\
line wings.  The data from all of our observing runs (each run
sampling only a small fraction of the orbit) indicate only small
variations on a 1-3\,d timescale for the blueshifted jet component and
larger variations only over months to years.  The variations in the
broad \halpha wings are best illustrated by the unusual profile at the
top of Fig. 2, with two additional broad peaks at
$\sim$$\pm$100\,\kms\ and wings extending up to $\pm$250\,\kms
. Similarly broad but much fainter wings are visible in all \halpha\
profiles, particularly on the blueshifted side. The nature of these
variable \halpha\ wings will be discussed in detail by Hamilton et
al.\ (2010) in terms of magnetospheric accretion using all available
profiles.

To estimate the full opening angle $\alpha$ of the jets and the angle,
$i_{jet}$, between the flow axis and the line of sight we only
consider the \halpha\ line profiles, because the [OI] profiles could
be blended with an unresolved low-velocity component (see above).  For
this purpose we assume that the jets can be approximated by a
homogeneously emitting cone. We will use v$_{rad}$ and FWHM values of
the \halpha\ profile from Dec. 16, 2003 to derive $i_{jet}$ and
$\alpha$ from eq. 5 of Mundt et al.\ (1990).  In this profile the two
peaks are at -30\,\kms\ (FWHM=39\,\kms) and +34\,\kms\
(FWHM=48\,\kms). If one assumes a jet velocity, v$_{jet}$, of 200
(100)\,\kms\ one derives $i_{jet}$=81$^{\circ}$(71$^{\circ}$) for
$\mid$v$_{rad}$$\mid$=32\,\kms\ and $\alpha$ values of
13$^{\circ}$(26$^{\circ}$) and 16$^{\circ}$(33$^{\circ}$) for the
redshifted and blueshifted jets, respectively.

\section{Possible jet launching models}

Before discussing possible models we like to remind the reader that
[OI]$\lambda$6300 emission in TTSs with small-scale jets is usually
extending over 10-30\,AU (e.g. Hirth et al.\ 1997).  We call this
spatial extend L$_{em}$, which is the length over which most of the
light from the jet is emitted. Since the KH\,15D jet has probably a
smaller mass flux than typical CTTSs L$_{em}$ in [OI] is more likely
10\,AU than 30\,AU (see also Sect. 5 for an independent estimate of
L$_{em}$).  For v$_{jet}$=100\,\kms\ it takes 180\,d to reach
10\,AU. If one adopts for KH\,15D such values, even a highly modulated
outflow with P$_{bin}$=48.4\,d would not cause detectable [OI] flux
variations, because the [OI] emission would be averaged out over
$\sim$3.7 periods.  Although \halpha\ is probably formed at smaller
length scales ($\sim$2-5.5\,AU; see Sect. 5) it is unlikely that
L$_{em}$(\halpha)/v$_{jet}$ is much shorter than the binary period and
therefore one would expect only a small period modulation of the
\halpha\ flux of the jet component.  These considerations are fully
supported by the small [OI] flux variations indicated by Fig. 1 and
the lack of periodic [OI] flux and profile variations in the CTTS
binaries DQ Tau and UZ Tau\,E (Basri et al.\ 1997, Huerta et al.\
2005).

Due to the high excentricity ($\sim$0.6) of the KH\,15D binary system
the two stars approach as close as $\sim$18\,R$_A$
(R$_A$=1.3\,R$_{\odot}$) during periastron, i.e.  near periastron any
individual circumstellar disks must be much smaller than this value
due to tidal truncation. We find it of great astrophysical interest
that such a close binary system can launch jets, and it remains as a
large theoretical challenge to model such a system, or even closer
systems like UZ Tau\,E.  One important guidance for any launching
model of the KH\,15D jets is the fact that the redshifted and
blueshifted jet components in \halpha\ and [OI] show a symmetric
velocity shift relative to v$_{sys}$=+18.6\,\kms .  In the following
we discuss two possible models which could explain this important
result.  In the first model the jets are launched from the innermost
part of the CBD.  A problem with this idea is the large inferred inner
radius of the CBD ($\sim$0.6\,AU; Herbst et al.\ 2008), since at this
radius the CBD is rotating with only v$_{rot}\sim$45\,\kms\ (P=145\,d,
3:1 inner Lindblad resonance). Magneto-hydrodynamic simulations
(e.g. Ouyed \& Pudritz 1997; Fendt 2006) for various magnetic field
strengths and magnetic field configurations have shown that jet
velocities of v$_{jet}\sim$1-1.5\,v$_{rot}$ are normal, although
v$_{jet}$$\sim$2\,v$_{rot}$ is not unusual.\ This means that
v$_{jet}$=100\,\kms\ is not implausible and we adopt this value for
the CBD launching model. According to the above $i_{jet}$ calculations
a v$_{jet}$ value of 100\,\kms\ would imply $i_{jet}$=71$^{\circ}$
which is roughly consistent with the inclination angle of the binary
system ($i_{bin}$=83-92$^{\circ}$) derived by Winn et al.\ (2006), in
the context of a specific geometric model of the occulting edge.
Better consistency might be achieved by allowing for a possible tilt
(by $\sim$10$^{\circ}$) between the binary plane and the innermost
part of the CBD, provided this tilt is oriented roughly along the line
of sight. According to this model binaries with shorter periods should
on average have higher jet speeds. This is in accordance with the [OI]
line profiles of UZ Tau\,E and DQ Tau, in which the edge of their blue
wing indicate v$_{jet}$ $\geq$150\kms\ and $\geq$120\kms ,
respectively (Huerta et al.\ 2005).

Although the CBD jet launching idea is attractive we also consider the
following second model, which produces higher jet speeds. This model
requires that each of the two binary stars launches a
magneto-hydrodynamically driven outflow with $\sim$200\,\kms\ from
each of the circumstellar disks and that each of these two outflows is
of similar mass flux and velocity. At some distance from the system
these two outflows merge to form a common jet. Whether this rather
speculative idea would work in the end requires detailed modeling.
One complication is the complex and unstable interaction of the
magnetospheres of the two outflows which should lead to reconnection
of the magnetic field lines.  Since in this model the line widths of
the jets should be correlated with the radial orbital motions of the
individual circumstellar disks (with v$_{rad}$ amplitudes of
$\sim$$\pm$50\,\kms ), it has to be shown, whether this model would
agree with the small measured velocity widths of the jet components
(FWHM=40-50\,\kms ). This model predicts a periodic modulation of the
outflow (see below), which would only be observable with an outflow
tracer emitting very close to the source (i.e., L$_{em}$/v$_{jet} \ll
P_{bin}$).

According to the models of G\"unther and Kley (2002) binary stars with
similar masses should have similar mass accretion rates
$\dot{M}_{acc}$ from their CBDs.  Due to the excentric orbit
($\epsilon$$\sim$0.6) $\dot{M}_{acc}$ should be highly modulated with
the binary period ($\dot{M}_{acc,max}$/$\dot{M}_{acc,min}$ $\sim$4-5)
and highest near periastron passage.  Since in TTSs
$\dot{M}_{outflow}$ is highly correlated with $\dot{M}_{acc}$, one
would expect for KH\,15D similar $\dot{M}_{outflow}$ values for the
two stars, which should in addition be highest near periastron. If,
quite in contrast to theoretical expectations, only one of the two
stars in the system would still drive the jets one would expect a
strongly modulated outflow with a higher fractions of the matter
ejected near periastron. Since at this phase each of the binary
members has a rather high v$_{rad}$ (~$\pm$50\,\kms\ relative to
v$_{sys}$) it would be hard to explain why the two jet components are
not symmetric about a different velocity, despite the fact that the
recorded jet emission traces gas ejected over several orbital
periods. These considerations make it rather unlikely than the jets of
KH\,15D are launched from only one of the two binary stars.

\section{Variations of  the redshifted jet  in \halpha }

The large variations in the ``eclipsed'' \halpha\ line profiles and
fluxes relative to the much more stable [OI] line profiles indicates
that these two lines cannot be formed in the same region.  Most
probably the [OI] emission traces more distant and tenuous
emission. This may mean that the [OI] emission is quenched in the
inner and denser \halpha\ emission region (i.e. N$_e$\,$>>$\,N$_{e,
  crit}\approx10^6$\,$cm^{-3}$) or that OI is largly ionized there.
Whether any emission component in our spectra is modulated with the
48.4\,d period cannot be answered with our data, because of
insufficient time sampling. A jet with 100\,\kms\ would reach a
distance of 2.7\,AU within 48.4\,d and, as already stated above, any
periodic modulation in \halpha\ would only be observable if L$_{em}$
in \halpha\ is much shorter than this value.  The shortest time scale
variations of the red jet component have been \ observed between
Dec.\,14 and 20, 2001. The large flux increase of this component by a
factor of 4 within 6\,d cannot be explained by the flux correction
uncertainties mentioned in Sect. 2. Perhaps they are related to the
unique ``accretion event'' which happened on Dec.\,20, 2001 (see
Hamilton et al.\ 2010). This could have produced sufficient UV
radiation to excite more distant neutral sections of the jet causing a
corresponding increase in \halpha\ on a small time scale.  The
increased UV flux might have also ionized OI causing the smaller [OI]
line flux observed on Dec.\,20, 2001.

We believe the large variations seen primarly in the redshifted
\halpha\ jet component are mainly a result of obscuration by the disk
combined with variation of L$_{em}$. Due to variations in the outflow
rate, or the above mentioned variations in UV excitation, L$_{em}$ for
both the blueshifted and redshifted jet will change with time. If the
redshifted jet is obscured by the circumstellar disk over a
considerable fraction, L$_{obsc}$, of its emitting section, then any
changes in L$_{em}$ will cause corresponding variations. If we assume
that the outer disk radius, R$_{disk}$, is at 5\,AU (see Chiang \&
Murray-Clay 2004) and that the disk plane is inclined towards the line
of sight by $\alpha$=15\degr\ then L$_{obsc}$=sin($\alpha$) R$_{disk}$
$\sim$1.3\,AU. If one assumes L$_{em}$(blue)=L$_{em}$(red) and the jet
emission rate per unit length is constant, one can derive from the
observed flux ratio of f(blue)/f(red)=f$_b$/f$_r$ the ratio
L$_{em}$/L$_{obsc}$=(f$_b$/f$_r$)/((f$_b$/f$_r$)-1), because
f$_b$/f$_r$=L$_{em}$(blue)/(L$_{em}$(red)-L$_{obsc}$). With the
measured f$_b$/f$_r$ values of $\sim$1.3-3.2 (and ignoring the unusual
profile of Dec.\ 20, 2001) one derives for \halpha\ L$_{em}$
$\sim$2-5.5\,AU. For [OI] the average ratio f$_b$/f$_r$ is 1.14. With
the above equation one derives L$_{em}$([OI])$\sim$10\,AU, in good
accordance with the values derived for other TTSs by
spectro-astrometric methods.



\acknowledgments We are grateful to Max Camenzind, Christian Fendt, and
\mbox{Hubert} Klahr for constructive discussions.

\clearpage



\begin{figure}
\epsscale{1.0}
\plotone{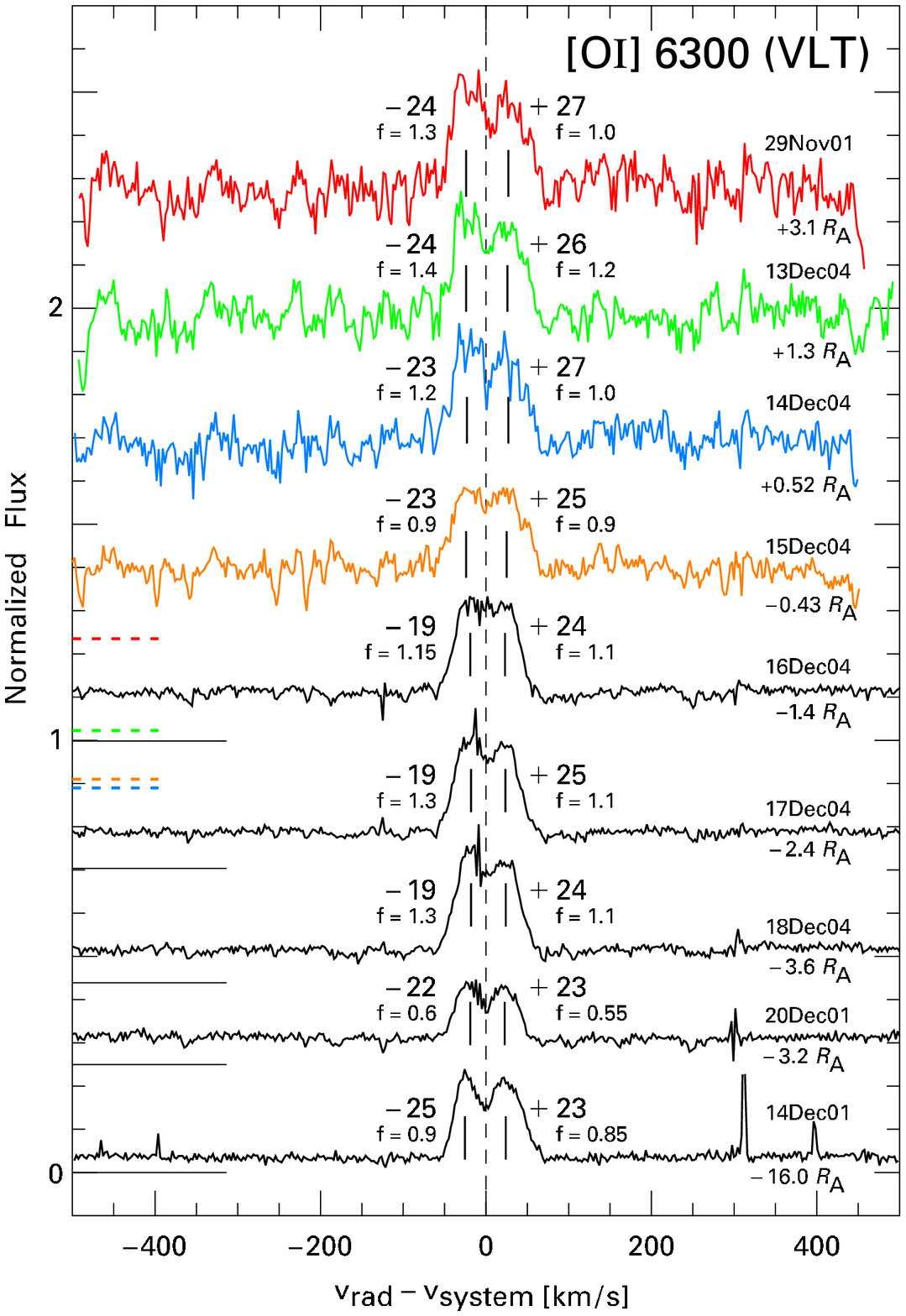}
\caption{ VLT \OI\ line profiles: the top four colored ones are taken
  out of eclipse (or at partial eclipse), while the five bottom ones
  are taken during eclipse.  In order to minimize profile changes due
  to the stellar absorption spectrum, we have subtracted from the top
  four profiles a properly scaled and broadened spectrum of the K5
  dwarf HD\,36003. The number indicated on the right hand side (below
  each profile) gives the distance between star A and the occulting
  edge in radii of star A (for details see Herbst et al.\ 2008).  The
  zero flux levels of the profiles are indicated on the left hand side
  (dashed colored lines for the four top profiles, black lines for the
  five bottom ones).  The velocities of the jet components are
  indicated in \kms\ and are based on Gaussian component fitting. The
  given f-values indicate the relative fluxes of the jet components,
  with the top right-hand component set to f=1.}
\end{figure}

\begin{figure}
\epsscale{0.78}
\plotone{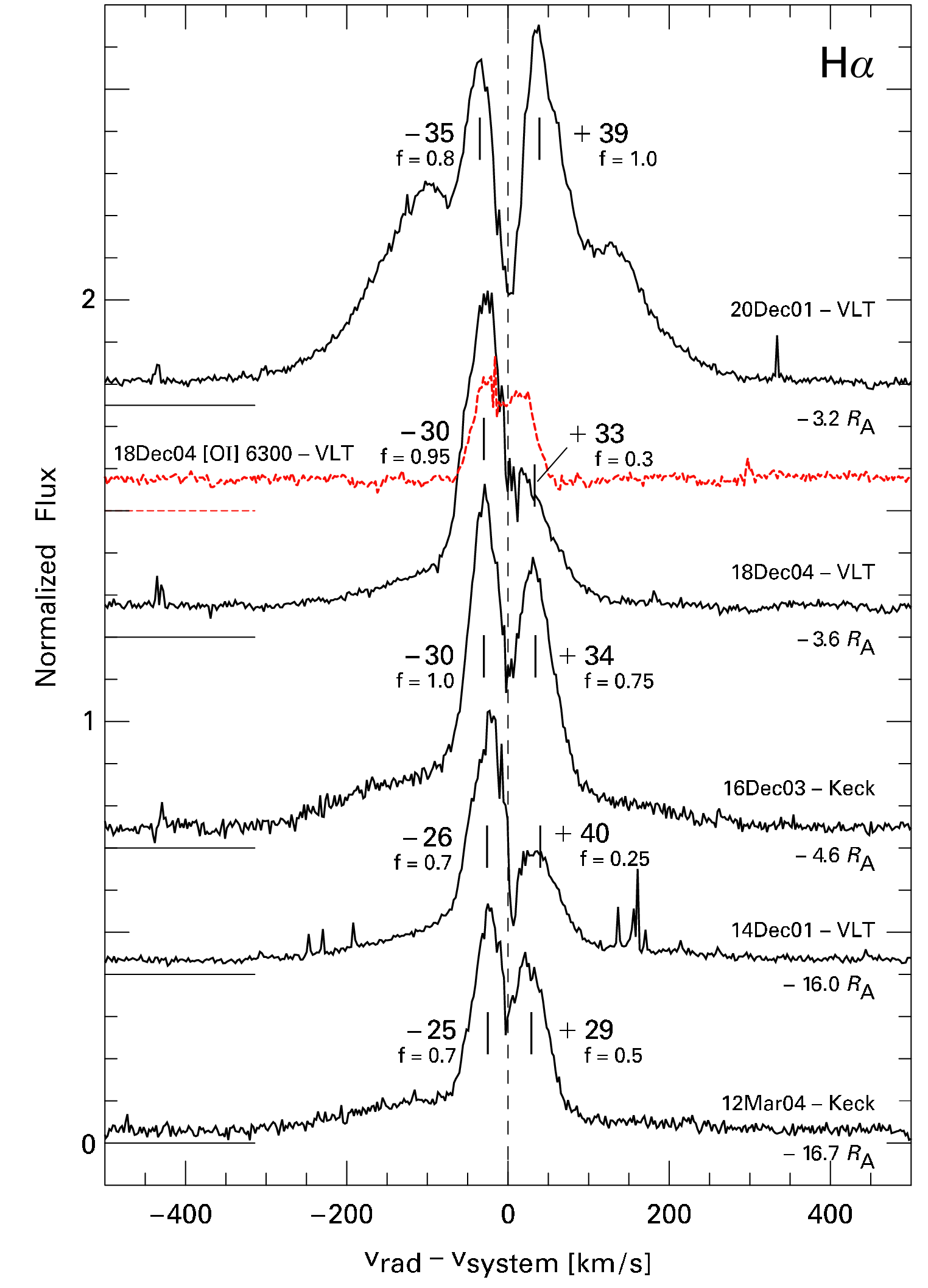}
\caption{Sample of \halpha\ line profiles obtained during eclipse at
  various observing runs at the VLT and KECK. The VLT \OI line profile
  from Dec. 18, 2004 is shown for comparison (dotted red line). For
  details see the caption of Fig. 1.}
\end{figure}





\begin{thebibliography}{}
\bibitem[Basri et al.(1997)]{1997AJ....114..781B} Basri, G., Johns-Krull, 
        C.~M., \& Mathieu, R.~D.\ 1997, \aj, 114, 781 
\bibitem[Chiang \& Murray-Clay(2004)]{2004ApJ...607..913C} Chiang, E.~I., 
        \& Murray-Clay, R.~A.\ 2004, \apj, 607, 913 
\bibitem[Eisl\"offel et al.(2000)]{2000prpl.conf..815E} Eisl\"offel, J., Mundt, 
       R., Ray, T.~P., \& Rodriguez, L.~F.\ 2000, Protostars and Planets IV, 815
\bibitem[Fendt(2006)]{2006ApJ...651..272F} Fendt, C.\ 2006, \apj, 651, 272
\bibitem[G{\"u}nther \& Kley(2002)]{2002A&A...387..550G} G{\"u}nther, R., 
      \& Kley, W.\ 2002, \aap, 387, 550 
\bibitem[Hamilton et al.(2003)]{2003ApJ...591L..45H} Hamilton, C.~M.,
        et al.\ 2003, \apjl, 591, L45 
\bibitem[Hamilton et al.(2005)]{2005AJ....130.1896H} Hamilton, C.~M., et 
              al.\ 2005, \aj, 130, 1896 
\bibitem[Hamilton et al.(2010)]{}  Hamilton, C.~M.,
        et al.\ 2010, in prep.
\bibitem[Hartigan et al.(1995)]{1995ApJ...452..736H} Hartigan, P., Edwards, 
        S., \& Ghandour, L.\ 1995, \apj, 452, 736 
\bibitem[Herbst et al.(2008)]{2008Natur.452..194H} Herbst, W.,  
        et al.\ 2008, \nat, 452, 194 
\bibitem[Hirth et  al.(1997)]{1997A&AS..126..437H} Hirth, G.~A., 
        Mundt, R., \& Solf, J.\ 1997, \aaps, 126, 437 
\bibitem[Huerta et al.(2005)]{2005AJ....129..985H} Huerta, M., Hartigan, P., 
        \& White, R.~J.\ 2005, \aj, 129, 985 
\bibitem[Leduc et al.(2009)]{} Leduc, K., et al.\ 2010, in prep.
\bibitem[Matt \& Pudritz(2008)]{2008ApJ...681..391M} 
        Matt, S., \& Pudritz, R.~E.\ 2008, \apj, 681, 391 
\bibitem[Mundt et al.(1990)]{1990A&A...232...37M} Mundt, R., 
        et al.\ 1990, \aap, 232, 37 
\bibitem[Ouyed \& Pudritz(1997)]{1997ApJ...482..712O} Ouyed, R., \& Pudritz, 
       R.~E.\ 1997, \apj, 482, 712 
\bibitem[Pudritz et al.(2007)]{2007prpl.conf..277P} Pudritz, R.~E., Ouyed, R., 
       Fendt, C., \& Brandenburg, A.\ 2007, Protostars and Planets V, 277 
\bibitem[Shang et al.(2007)]{2007prpl.conf..261S} Shang, H., Li, Z.-Y., 
        \& Hirano, N.\ 2007, Protostars and Planets V, 261 
\bibitem[Winn et al.(2004)]{2004ApJ...603L..45W} Winn, J.~N.,
         et al.\ 2004, \apjl, 603, L45         
\bibitem[Winn et al.(2006)]{2006ApJ...644..510W} Winn, J.~N.,  
        et al.\ 2006, \apj, 644, 510 

\end{thebibliography}
\end{document}